\documentclass[aps,twocolumn,showkeys,amsmath,amssymb,superscriptaddress,showpacs]{revtex4}
\usepackage{graphicx}
\usepackage{color}
\usepackage{makecell}
\newcommand{\laco}{LaCoO$_3$}

\newcommand{\ndco}{NdCoO$_3$}
\newcommand{\tbco}{TbCoO$_3$}

\newcommand{\dyco}{DyCoO$_3$}
\newcommand{\smco}{SmCoO$_3$}

\newcommand{\st}{$\rm ^o$}
\newcommand{\stc}{$\rm ^oC$}

\newcommand{\neel}{N\'{e}el}
\newcommand{\emumoloe}{emu$\cdot$mol$^{-1}\cdot$Oe$^{-1}$}
\newcommand{\teLSiii}{$t_{2g}^6e_g^0$}

\newcommand{\teHSiii}{$t_{2g}^4e_g^2$}
\newcommand{\figwp}{0.95}
\newcommand{\figws}{1.00}
\newcommand{\tabws}{0.70}
\begin{document}
\title{Non-collinear magnetic structures of TbCoO$_3$ and DyCoO$_3$}

\author{K. Kn\'{\i}\v{z}ek $^{1}$}
\email[corresponding author:]{knizek@fzu.cz}
\author{Z. Jir\'{a}k $^{1}$      }
\author{P. Nov\'{a}k $^{1}$      }
\author{Clarina R. dela Cruz $^{2}$}
\affiliation{
 $^{1}$ Institute of Physics ASCR, Cukrovarnick\'a 10, 162 00 Prague 6, Czech Republic }
\affiliation{
 $^{2}$ Neutron Scattering Science Division, ORNL, Oak Ridge, Tennessee 37831, United States }
\begin{abstract}
The orthoperovskites TbCoO$_3$ and DyCoO$_3$ with Co$^{3+}$ in a non-magnetic low-spin state
have been investigated by neutron diffraction down to 0.25~K. Magnetic ordering is evidenced
below $T_N=3.3$~K and 3.6~K, respectively, and the ordered arrangements are of canted type,
A$_x$G$_y$ for TbCoO$_3$ and G$_x$A$_y$ for DyCoO$_3$ in Bertaut's notation. The experiments are
confronted with the first-principle calculations of the crystal field and magnetism of Tb$^{3+}$
and Dy$^{3+}$ ions, located in the $Pbnm$ structure on sites of $C_s$ point symmetry. Both these
ions exhibit an Ising behavior, which originates in the lowest energy levels, in particular in
accidental doublet of non-Kramers Tb$^{3+}$ ($4f^8$ configuration) and in ground Kramers doublet
of Dy$^{3+}$ ($4f^9$) and it is the actual reason for the non-collinear AFM structures. Very
good agreement between the experiment and theory is found. For comparison, calculations of  the
crystal field and magnetism for other systems with Kramers ions, NdCoO$_3$ and SmCoO$_3$, are
also included.
\end{abstract}
\maketitle
\section{Introduction}

The perovskite cobaltites of \laco\ type are systems with closeness in energy of different local
state of the octahedrally coordinated cobalt ions. The ground state is generally based on
non-magnetic low spin state of Co$^{3+}$ (LS, $S=0$, \teLSiii), and the paramagnetic high spin
Co$^{3+}$ (HS, $S=2$, \teHSiii) species are populated by thermal excitation, which process becomes
observable for \laco\ above $\sim 40$~K and achieves the highest rate at $T_{magn}=70$~K as
documented by a maximum of the specific heat excess and anomalous expansion. The resulting
non-uniform phase is characterized by HS/LS nearest neighbor correlations
\cite{RefGoodenough1958JPCS6_287,RefKnizek2009PRB79_014430GGA}. A further change is observed at
about 530~K where the correlations are melted by thermal agitation, which is accompanied with a
drop of electrical resistivity reminding the insulator-metal transition. In related systems with
smaller rare-earth or yttrium ions, the magnetic transition shifts to higher temperatures and
finally merges with the electrical transition, in particular for \tbco\ and \dyco\ at
$T_{magn}=735$~K and 740~K, respectively \cite{RefKnizek2005EPJB47_213}. Such increased stability
of LS Co$^{3+}$ means that the low-temperature magnetic properties are governed by Tb$^{3+}$ and
Dy$^{3+}$ local moments and depend on the crystal field splitting of respective $4f^8$ and $4f^9$
levels, as well as on the rare-earth intersite interactions.

Early studies have shown that \tbco\ and \dyco\ undergo a non-collinear antiferromagnetic (AFM)
ordering below $T_N=3.3$~K and 3.6~K due to interactions that are for the most part of a classical
dipole-dipole nature \cite{RefMareschal1968JAP39_1364,RefKappatsch1970JPHYSFRAN31_369}. The
present study revisits both systems. The crystal and magnetic structures are refined based on the
high-resolution powder neutron diffraction, and the electronic levels of rare-earth ions are
determined together with their magnetic characteristics, using a novel \textit{ab-initio}
approach. It appears that the observed magnetic arrangements reflect the Ising nature of
rare-earths moments associated with the lowest energy levels, namely the ground Kramers doublet of
Dy$^{3+}$ and accidental doublet of non-Kramers Tb$^{3+}$.

\section{Experiment and Calculations}

Samples \tbco\ and \dyco\ were prepared by a solid state reaction from stoichiometric amounts of
Co$_2$O$_3$ and respective oxides Tb$_2$O$_3$ and Dy$_2$O$_3$. The precursor powders were mixed,
pressed in the form of pellets and sintered at 1200\stc\ for 100~hours under air. The product was
checked for phase purity and its structural and physical properties were extensively probed. Both
compounds were identified as distorted perovskite structure of the orthorhombic $Pbnm$ symmetry.

The powder neutron diffraction was performed on diffractometer Hb2a at Oak Ridge National
Laboratory. The scans were recorded at selected temperatures between the 0.25 and 150~K. Two
crystal monochromators (Ge113 and Ge115) were used, providing neutron wavelengths
$\lambda=2.408$~\AA\ and 1.537~\AA, respectively. Data were collected between 8\st\ and 126\st\ of
2$\theta$ with the step of 0.08\st. To overcome the problem with rather high absorption of
dysprosium, the neutrons path length through the \dyco\ sample was minimized by use of an annular
container, where the powder was placed between two concentric aluminium cylinders. Structural
refinements were done by Rietveld profile analysis using program FULLPROF (Version
5.30-Mar2012-ILL JRC).

The determination of the Tb$^{3+}$ and Dy$^{3+}$ electronic structure was performed in two steps.
First, the crystal field parameters were calculated using a novel method based on the
first-principles band structure and Wannier projection \cite{RefNovak2013PRB87_205139}. The model
depends on a single parameter $\Delta$, which adjust relative position of the $4f$ and oxygen
energy levels. In the present calculations $\Delta$ was fixed at 8.2~eV, which value yields very
good agreement between the calculated  and experimental values of the multiplet splittings of
Pr$^{3+}$, Nd$^{3+}$, Tb$^{3+}$ and Er$^{3+}$ in the orthoperovskites
\cite{RefNovak2013PRB87_205139,RefNovak2013JPCM}. Then the local Hamiltonian operating on the $4f$
states in a determinantal basis of one-electron wavefunctions was constructed, including the
electron-electron repulsion, spin-orbit coupling, crystal field and Zeeman interaction terms. The
eigenvalue problem was solved for different orientation and strength of the applied field, in
which not only the crystal field splitting of the multiplets, but also their magnetic
characteristics (anisotropic $g$-factors and/or van Vleck susceptibilities) were determined. These
calculations were done with the modified 'lanthanide' program \cite{RefEdvardsson2001CPC133_396}.

\section{Results}

\subsection{Crystal structure}

\begin{table*}
\caption{
The crystallographic data summary for \tbco\ and \dyco\ at selected temperatures.
Space group $Pbnm$.
}
\centering \setcellgapes{2.0pt} \makegapedcells
\begin{tabular*}{\textwidth}{@{\extracolsep{\fill}} l|rrr|rrr}
\hline \hline
                        & \multicolumn{3}{c|}{\tbco}           & \multicolumn{3}{c}{\dyco}            \\
\hline
T (K)                   & 0.25       & 5.5        & 150        & 0.25       & 5.5        & 150         \\
\hline
a (\AA)                 & 5.2034(3)  & 5.2034(3)  & 5.1995(2)  & 5.1644(3)   & 5.1650(3)   & 5.1655(3)   \\
b (\AA)                 & 5.3890(3)  & 5.3898(3)  & 5.3945(2)  & 5.4165(3)   & 5.4161(3)   & 5.4143(4)   \\
c (\AA)                 & 7.4050(4)  & 7.4052(4)  & 7.4102(3)  & 7.3806(4)   & 7.3813(4)   & 7.3866(5)   \\
V (\AA$^3$)             & 207.64(2)  & 207.68(2)  & 207.84(2)  & 206.46(2)   & 206.48(2)   & 206.59(2)   \\
\hline
\multicolumn{7}{c}{Atom coordinates: TbDy $4c$(x,y,1/4), Co $4b$(1/2,0,0), O1 $4c$(x,y,1/4), O2 $8d$(x,y,z).}            \\
\hline
x,TbDy                  & -0.0124(7) & -0.0121(6) & -0.0134(4) & -0.0119(9)  & -0.0116(8)  & -0.0114(8)  \\
y,TbDy                  & 0.0633(4)  & 0.0597(4)  & 0.0608(3)  & 0.0618(6)   & 0.0649(4)   & 0.0641(4)   \\
x,O1                    & 0.0924(9)  & 0.0919(7)  & 0.0921(5)  & 0.0905(14)  & 0.0952(13)  & 0.0940(14)  \\
y,O1                    & 0.4779(8)  & 0.4772(6)  & 0.4776(4)  & 0.4942(15)  & 0.4824(11)  & 0.4844(12)  \\
x,O2                    & -0.3008(6) & -0.3000(4) & -0.2994(3) & -0.3162(14) & -0.3128(10) & -0.3132(11) \\
y,O2                    & 0.2970(6)  & 0.2976(4)  & 0.2971(3)  & 0.3039(13)  & 0.3046(10)  & 0.3051(10)  \\
z,O2                    & 0.0446(4)  & 0.0461(3)  & 0.0454(2)  & 0.0473(8)   & 0.0487(6)   & 0.0483(6)   \\
\hline
\multicolumn{7}{c}{Selected bond distances and angles.}             \\
\hline
Co-O1 (\AA)$\times 2$   & 1.916(2)   & 1.916(1)   & 1.917(1)   & 1.904(3)    & 1.912(2)    & 1.911(2)    \\
Co-O2 (\AA)$\times 2$   & 1.935(4)   & 1.935(3)   & 1.932(2)   & 1.932(8)    & 1.945(6)    & 1.946(6)    \\
Co-O2 (\AA)$\times 2$   & 1.938(4)   & 1.942(3)   & 1.942(2)   & 1.979(8)    & 1.965(6)    & 1.964(6)    \\
Co-O1-Co(\st)$\times 1$ & 150.0(4)   & 150.1(3)   & 150.1(2)   & 151.5(8)    & 149.6(6)    & 150.1(6)    \\
Co-O2-Co(\st)$\times 2$ & 150.5(4)   & 150.1(3)   & 150.5(2)   & 146.2(8)    & 146.3(6)    & 146.2(6)    \\
\hline \hline
\end{tabular*}
\label{tab:struct}
\end{table*}

For \tbco\ and \dyco\ the single orthorhombic $Pbnm$ structure is observed over the whole
experimental temperature range. For three selected temperatures the complete data including atomic
coordinates, Co-O bonding lengths and Co-O-Co angles are presented in Table~\ref{tab:struct}. The
relation between lattice parameters, $b > c/\sqrt 2 > a$ is typical for perovskites ABO$_3$ with
smaller A cations (and smaller tolerance factor), for which the tilting of CoO$_6$ octahedra is
the dominant source of the orthorhombic distortion. The extent of octahedral tilting is quantified
by the average bond angle Co-O-Co, which is close to 150\st\ for both  the compounds. The
rare-earth cations are located on mirror plane of the $Pbnm$ structure, on the site of low
symmetry ($C_s$ point group), which is coordinated by twelve oxygen atoms. The RE-O distances
could be divided into a group of eight short (bonding) lengths within a range $2.2-2.6$~\AA, and a
group of four long distances $3.0-3.4$~\AA\ that should be considered as non-bonding ones since
they increase with decreasing A size.

\subsection{Electronic levels of Tb$^{3+}$ and Dy$^{3+}$}

Tb$^{3+}$ in electronic configuration $4f^8$ is a non-Kramers ion. The lowest lying free-ion term
is $^7F_{6}$ ($L=3$, $S=3$, $J=6$), and in the solid it is split by crystal field effects. In
$Pbnm$ perovskites with the rare-earth sites of low symmetry $C_s$, the degeneracy is completely
removed, yielding 13 singlets. The analysis of optical transitions available for TbAlO$_3$ of the
same $Pbnm$ symmetry show that the ground and first excited states of Tb$^{3+}$ ion are formed of
90\% by two conjugate $J=6$ wavefunctions, $|6,+6\rangle+|6,-6\rangle$ and
$|6,+6\rangle-|6,-6\rangle$, and their eigenenergies differ by only 0.025~meV, representing a
quasi-doublet \cite{RefGruber2008JLUM128_1271}. This specific kind of accidental degeneracy has
important consequences. Firstly, a relatively modest magnetic field of external or molecular
nature will mix the eigenstates into a form of two pseudospins with the main weight of
$|6,+6\rangle$ and $|6,-6\rangle$ ionic states, respectively. This results in large magnetic
moment of Tb$^{3+}$ ions of about 8.4~$\mu_B$. Secondly, these moments have essentially an
Ising-like character, which is a source of large local anisotropy.

In the case of Dy$^{3+}$ in electronic configuration $4f^9$, the lowest lying free-ion term is
$^6H_{15/2}$ ($L=5$, $S=5/2$, $J=15/2$), and it is split by the low-symmetry crystal field to 8
Kramers doublets. The ground doublet is spanned by two pseudospins with dominant contribution of
$|15/2,+15/2\rangle$ and $|15/2,-15/2\rangle$ ionic states, pointing also to the Ising character.

The results of present first-principle approach are summarized in Appendix, where the calculated
single-electron CFP for non-Kramers Tb$^{3+}$ and Kramers Dy$^{3+}$ are displayed including two
other ions possessing Kramers degeneracy, Nd$^{3+}$ ($4f^3$) and Sm$^{3+}$ ($4f^5$) (see
Table~\ref{tab:cfp}). Using these parameters in the 'lanthanide' program, the splitting of $4f^n$
levels for each rare-earth ion is determined. It is seen below in Table~\ref{tab:tbco_e_g} that
the two lowest singlets for Tb$^{3+}$ in \tbco\ are separated in energy by only 0.002~meV. Next
two singlets are situated at much larger energy, 22.9 and 23.1 meV. For Dy$^{3+}$ all the states
are true doublets and the first excited doublet is situated at energy of 29.8~meV above the ground
doublet. Similar separation between the ground and excited doublet is found for Sm$^{3+}$, 29.6
meV, while a somewhat lower separation 13.2~meV is obtained for Nd$^{3+}$ (see also
\cite{RefNovak2013JPCM}).

\begin{figure}
\includegraphics[width=\figwp\columnwidth,viewport=0 190 582 819,clip]{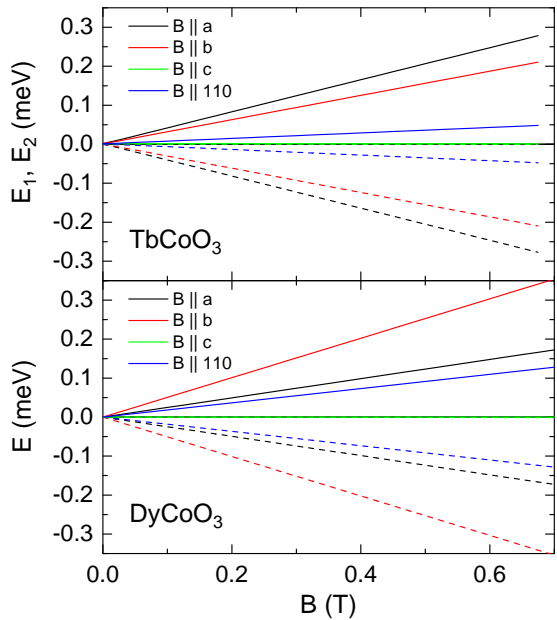}
\caption{Zeeman splitting of the \tbco\ low-energy pseudodoublet and the \dyco\ ground doublet.}
\label{fig:Zeeman}
\end{figure}

The application of external magnetic field brings two effects on the rare-earth electronic levels
(see \textit{e.g.} \cite{RefNovak2013JPCM}). One is a linear splitting of the doublets that
defines the relevant magnetic moments and represents Zeeman energy. Except very low fields, the
splitting of the  Tb$^{3+}$ pseudodoublet is also linear, as illustrated for several orientations
of applied field in upper panel of Fig.~\ref{fig:Zeeman}. The corresponding $g$-factors are
strongly anisotropic, they are in fact of Ising character with principal components $g_x=17.78$,
$g_y\sim0$, $g_z\sim0$, where the local $z$-axis is defined by symmetry along the perovskite $c$
axis and the Ising $x$-axis in $c$-plane makes an angle $\alpha_g =\pm 37.1$\st\ with the
orthorhombic $a$ axis. (Here $\pm$ refers to two inequivalent rare-earth sites in the structure.)

An analogous calculation for the ground doublet of Dy$^{3+}$ (see lower panel of
Fig.~\ref{fig:Zeeman}) gives principal components $g_x=19.44$, $g_y\sim0$, $g_z\sim0$ and
orientation of the Ising axis in $c$-plane under the angle $\alpha_g =\pm 64.0$\st\ to the
$a$-axis.

The $g$-factors for Sm$^{3+}$ and Nd$^{3+}$ are much less anisotropic with principal in-plane
components $g_x=0.703$, $g_y=0.588$, out-of-plane component $g_z=0.322$ and orientation of the
$x$-axis under the angle $\alpha_g =\pm 33.2$\st\ for \smco. The respective values for \ndco\ are
$g_x=2.818$, $g_y=1.228$, $g_z=3.015$ and  angle $\alpha_g =\pm 62.3$\st.

The second effect, visible in high fields only, is a small downward shift of the doublet gravity
center that varies quadratically with the field magnitude. This energy decrease arises from the
mixing of ground singlets or doublets with higher lying states and is at the root of van Vleck
paramagnetism. In systems with non-Kramers ions, this term is generally manifested at low
temperatures as a practically constant susceptibility, independent up to very high fields. The
well known example is the paramagnetic contribution of Pr$^{3+}$ \cite{RefThomas1999JAP85_5384}.
Similar contribution exists also for Kramers ions, where van Vleck susceptibility represents a
small addition to dominant, strongly temperature dependent Curie susceptibility that arises due to
thermal redistribution within the split levels of ground doublet. In particular for \dyco, the
principal values of the van Vleck susceptibility tensor associated with Dy$^{3+}$ are calculated
to $\chi_\xi$=0.0143, $\chi_\eta \sim$0 for the in-plane components, $\chi_\zeta=0.0526$~$\mu_B$/T
for the out-of-plane component and the orientation of $\xi$ axis is defined by $\alpha_{vV} =\pm
22.2$\st\ with respect to the $a$-axis. Let us note that there is also van Vleck contribution of
LS Co$^{3+}$, which is of the order $\sim0.0004$~$\mu_B$/T ($\sim0.0002$~\emumoloe) and brings
thus little effect.

\subsection{Magnetic ordering}

\begin{figure}
\includegraphics[width=\columnwidth,viewport=50 240 770 800,clip]{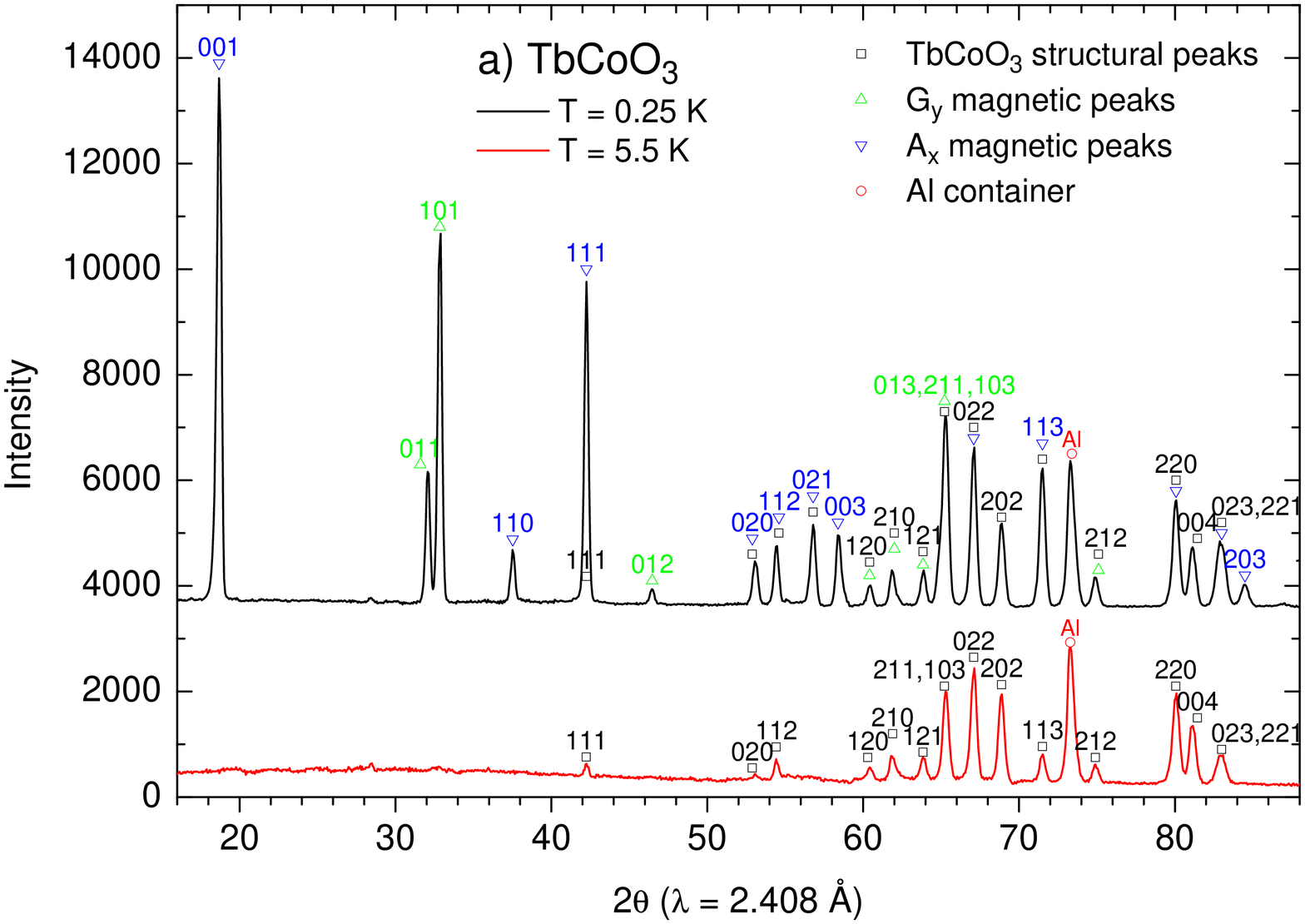}
\includegraphics[width=\columnwidth,viewport=50 210 770 800,clip]{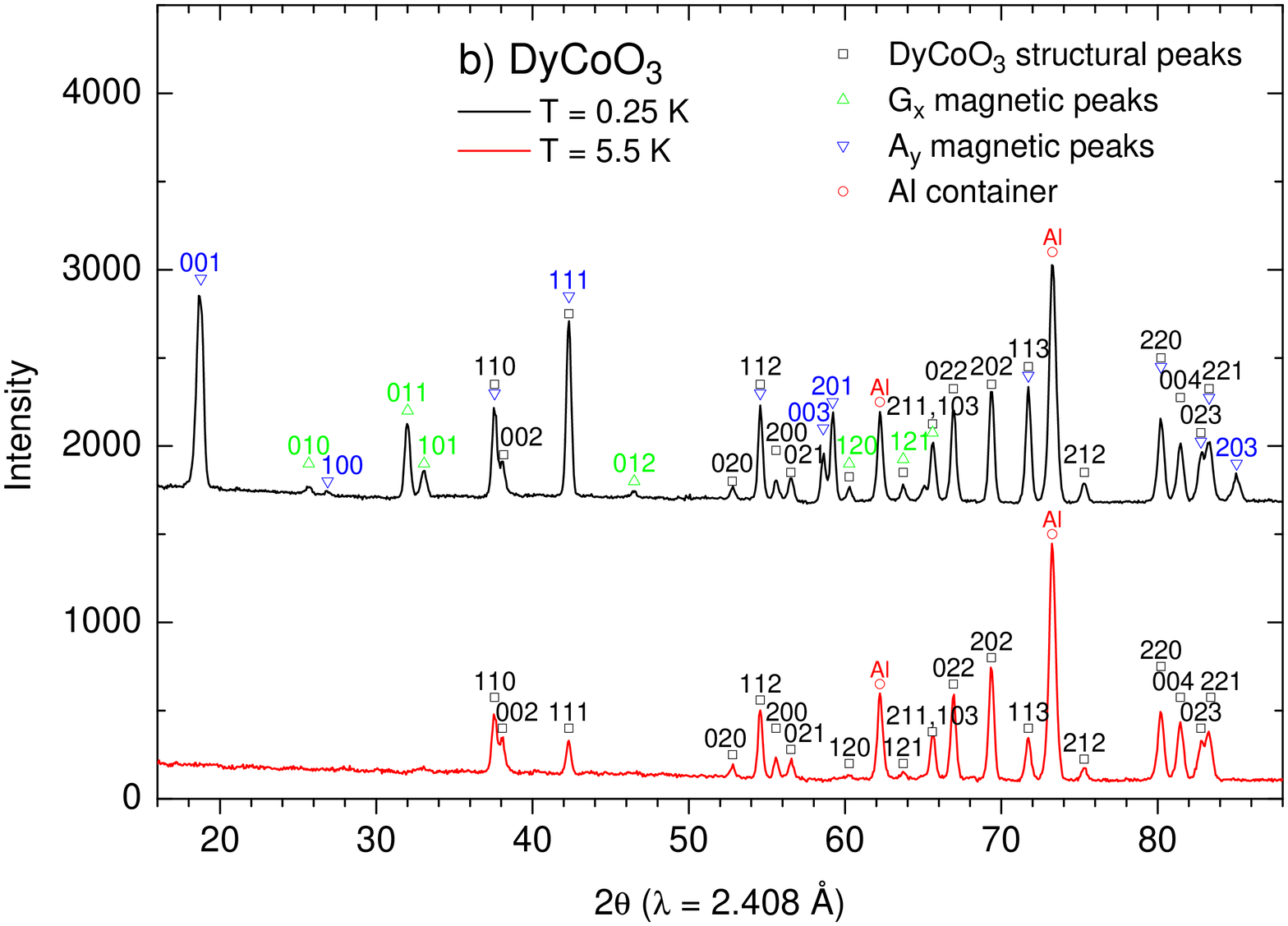}
\caption{Neutron diffraction patterns of \tbco\ and \dyco\ at 0.25 and 5.5~K ($\lambda=2.408$~\AA).}
\label{fig:NDdif}
\end{figure}

The neutron diffraction data at the lowest temperature ($T=0.25$~K) and above $T_N$ ($T=5.5$~K)
are displayed in Fig.~\ref{fig:NDdif}. Presence of AFM ordering is manifested in appearance of
magnetic peaks below $T_N$ at positions $hkl$ with $h+k=2n+1$, in particular 100+010 and 101+011.
These peaks are indicative for the A- and G-types of AFM arrangement in the sample, depending
whether $l=2n$ or $2n+1$. Let us note that the presence of both components indicates a canting,
which is an inevitable consequence of the Ising character of Tb$^{3+}$ and Dy$^{3+}$ moments, and
the inclination of easy axes $\pm\alpha$ in the \textit{ab}-plane of the orthoperovskite structure
\cite{RefGruber2008JLUM128_1271}.

\begin{figure}
\includegraphics[width=\figwp\columnwidth,viewport=0 400 582 819,clip]{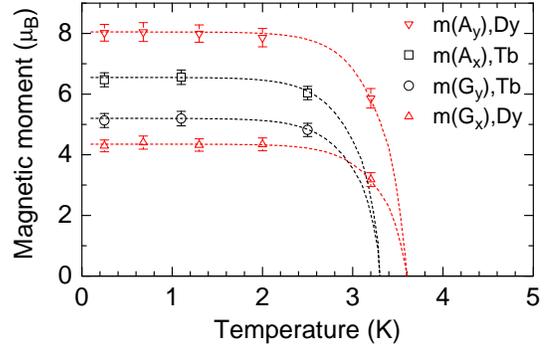}
\caption{The AFM ordered moments deduced from neutron diffraction patterns of \tbco\ and \dyco.
The dashed lines are guides to the eyes. } \label{fig:NDmag}
\end{figure}

The results of the Rietveld refinement of the rare-earth moments are presented in
Fig.~\ref{fig:NDmag}. The estimated \neel\ temperatures of the AFM ordering $T_N=3.3$~K and 3.6~K
for \tbco\ and \dyco, respectively, are in agreement with previous studies
\cite{RefMareschal1968JAP39_1364,RefKappatsch1970JPHYSFRAN31_369,RefMunoz2012EJICH2012_5825}.
The ordered moments are oriented in the $a,b$-plane of the orthoperovskite structure and reach
$m(A_x) = 6.5 \mu_B$ and $m(G_y) = 5.1 \mu_B$ for \tbco\ and $m(G_x) = 4.3 \mu_B$ and $m(A_y) =
8.0 \mu_B$ for \dyco.

\section{Discussion}

\begin{figure}
\includegraphics[width=\figws\columnwidth,viewport=50 500 500 800,clip]{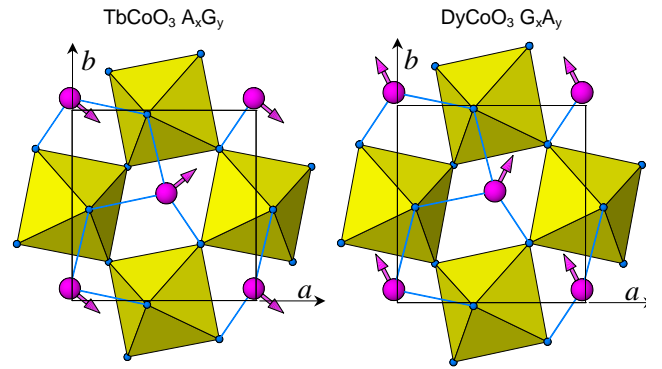}
\caption{Schematic view of octahedral tilts and the magnetic structures of \tbco\ and \dyco.
The CoO$_6$ octahedrons are centered at the $c=0$ level; the displayed rare-earth
moments are located at $c=1/4$ while those at $c=3/4$ are oppositely oriented.}
\label{fig:struct}
\end{figure}

Based on the present study, \neel\ temperatures of the AFM ordering in \tbco\ and \dyco\ is
estimated to $T_N=3.3$~K and 3.6~K, and the ordering is of A$_x$G$_y$ and G$_x$A$_y$ type,
respectively. This is in agreement with earlier findings on these cobaltites
\cite{RefKappatsch1970JPHYSFRAN31_369,RefMunoz2012EJICH2012_5825} and is also close to AFM
temperatures in analogous aluminates \cite{RefBidaux1968JDEPHYS29_220,RefHolmes1968JAP39_1373}.
The arrangements in the $a,b$-plane of the orthoperovskite structure are depicted in
Fig.~\ref{fig:struct} and the coupling along the $c$-axis is purely AFM. The ordered moments for
\tbco\ have a magnitude of $m=8.3 \mu_B$ and are inclined from the orthoperovskite $a$-axis to
angle $\pm38$\st. For \dyco, the magnitudes of moments is $m=9.1 \mu_B$ and their inclination is
$\pm62$\st.

The experimental magnitudes and directions of the magnetic moments are confronted with calculated
values in Table~\ref{tab:expcalc}. Calculated magnitudes of the moments are by 7 and 6 \% larger
than the experimental ones in \tbco\ and \dyco, respectively.
Taking into account that theory contains no parameter to fit the experiment the agreement may be
regarded as good for the magnitudes of the moments and excellent for their directions. As pointed
out by Bidaux and Meriel for DyAlO$_3$ \cite{RefBidaux1968JDEPHYS29_220}, there are two variants
depending in which sense the moments could be inclined, which cannot be distinguished in neutron
diffraction experiment. On the other hand, theory allows an unambiguous assignment of the sign of
angle $\alpha_g$ to the $RE$ crystal site. Regarding the agreement of the experimentally
determined angles and the orientation of Ising axes in our calculations, we may state that, with
respect to CoO$_6$ tilts, the actual orientation is as shown in Fig.~\ref{fig:struct}.

As mentioned above, the origin of observed AFM arrangement, $A_xG_y$ for \tbco\ and $G_xA_y$ for
\dyco\ in Bertaut's notation, can be associated with dipolar interactions, which dominant role can
be understood in view of large magnitude of the Tb$^{3+}$ and Dy$^{3+}$ long-range ordered
moments, $\sim 8$ and 9~$\mu_B$, respectively. There are, however, strong indications that the
superexchange interactions of purely spin nature must be also acting and may eventually decide on
final magnetic arrangement in the rare-earth cobaltites, aluminates or gallates. In particular,
AFM ordering at $T_N=1.5$~K is stabilized for \smco, though the spin and orbital components of the
Sm$^{3+}$  moment nearly cancel ($L=5$, $S=5/2$, $J=L-S=5/2$), and dipolar interactions are thus
negligible.

\begin{table}
\caption{Comparison of the rare-earth magnetic moment ($m_{RE}$)
and its inclination from the orthoperovskite $a$-axis ($\alpha_g$)
obtained by neutron diffraction refinement (ND) and electronic structure calculation (calc.)
for the reference position ($\sim0$, $\sim0$, 1/4).
}
\centering \setcellgapes{3.5pt} \makegapedcells
\begin{tabular*}{\columnwidth}{@{\extracolsep{\fill}} l|rr|rr}
\hline \hline
                  & \multicolumn{2}{c|}{\tbco} & \multicolumn{2}{c}{\dyco} \\
\hline
                  & ND          & calc.         & ND          & calc.      \\
\hline
$\alpha_g$(\st)   & $\pm 38$(1) & $-37.1$       & $\pm 62$(1) & $-64.0$    \\
$m_{RE}$($\mu_B$) & 8.3(2)      & 8.9           & 9.1(2)      & 9.7        \\
\hline \hline
\end{tabular*}
\label{tab:expcalc}
\end{table}


\textbf{Acknowledgments}. This work was supported by Project No.~P204/11/0713 of the Grant Agency
of the Czech Republic. We acknowledge the Oak Ridge National Laboratory (Tennessee, United States)
for providing access to the neutron beams and for all technical support during the experiments.

\newpage

\appendix
\section{Electron states and magnetism of lanthanide ions in orthocobaltites}

The crystal field parameters, obtained by first-principle calculations with a use of experimental
crystallographic data for \ndco\ \cite{RefKnizek2009PRB79_134103ND}, \smco\
\cite{RefPerezCacho2000JSSC150_145}, \tbco\ and \dyco\ (see Table~\ref{tab:struct}), are
summarized in Table~\ref{tab:cfp}. The energies of Kramers doublets and respective $\hat{g}$ and
$\hat{\chi}^{vV}$ tensor components for Nd$^{3+}$, Sm$^{3+}$ and Dy$^{3+}$ are presented in
Tables~\ref{tab:ndco_e_g}, \ref{tab:smco_e_g} and \ref{tab:dyco_e_g}. The diagonal tensor
components refer to axes $a$, $b$ and $c$ of the orthorhombic space group $Pbnm$, while the index
$\omega$ stands for the field orientation along 110.

The energies of thirteen levels of the $^7F_6$ multiplet of the non-Kramers Tb$^{3+}$ are given in
Table~\ref{tab:tbco_e_g}. It appears that there are five pairs of states with close energy
possessing a quasi-doublet character, \textit{i.e.} allowing sizeable Zeeman splitting when the
strength of external field exceeds the energy gap. The remaining three singlets are non-magnetic
and their response to external field is of the van Vleck character.

\begin{table*}
\caption{Nonzero parameters of the crystal field (in meV) in four compounds studied.
They hold for the rare-earth position close to ($\sim0$, $\sim0$, 1/4)}
\centering \setcellgapes{3.5pt} \makegapedcells
\begin{tabular*}{\tabws\textwidth}{@{\extracolsep{\fill}} cccccc}
\hline \hline
k & q & NdCoO$_3$ & SmCoO$_3$ & TbCoO$_3$ & DyCoO$_3$ \\
\hline
2 & 0 & $28.27 $ & $-37.47 $ & $-21.73 $ & $-65.93 $ \\
2 & 2 & $6.47 +94.60i$ & $8.31 +104.64i$ & $-9.42 +62.64i$ & $-63.09 +121.03i$ \\
4 & 0 & $-60.52 $ & $-58.65 $ & $-35.01 $ & $-40.22 $ \\
4 & 2 & $-9.92 +120.83i$ & $-40.05 +83.35i$ & $-22.50 +62.48i$ & $-59.49 +52.97i$ \\
4 & 4 & $46.45 -64.44i$ & $33.15 -93.32i$ & $19.07 -93.94i$ & $12.53 -128.77i$ \\
6 & 0 & $-121.36 $ & $-66.87 $ & $-54.23 $ & $-41.20 $ \\
6 & 2 & $8.78 +63.53i$ & $11.67 +33.29i$ & $14.34 +25.77i$ & $11.46 +22.01i$ \\
6 & 4 & $-195.90 +6.00i$ & $-136.29 -0.34i$ & $-115.29 -6.27i$ & $-92.65 -4.13i$ \\
6 & 6 & $11.63 +0.48i$ & $11.68 -3.65i$ & $11.18 +1.53i$ & $2.55 +2.49i$ \\
\hline \hline
\end{tabular*}
\label{tab:cfp}
\end{table*}

\begin{table*}
\caption{
Nd$^{3+}$ ion in NdCoO$_3$. Energy of five Kramers doublets originating from $^4I_{9/2}$ multiplet,
$\hat{g}$ and $\hat{\chi}^{vV}$ tensor components along the orthorhombic axes and $\omega$ direction.
Energy  $\varepsilon(0)$ is in meV, $\hat{\chi}^{vV}$ is units of $\mu_B$/T. }
\centering \setcellgapes{3.5pt} \makegapedcells
\begin{tabular*}{\textwidth}{@{\extracolsep{\fill}} cccccccccc}
\hline
\hline
doublet & $\varepsilon(0)$ &$g_{aa}$& $g_{bb}$& $g_{cc}$& $g_{\omega}$&$\chi^{vV}_{aa}$& $\chi^{vV}_{bb}$&$\chi^{vV}_{cc}$&$\chi^{vV}_{\omega}$ \\
\hline
 1 & 0.00 &1.701 &2.560 &3.015 &1.442 &0.0149 &0.0106 &0.0118 &0.0024 \\
 2 &13.19 &1.773 &2.208 &2.432 &0.813 & -0.0072 & -0.0020 &0.0040 &0.0054 \\
 3 &25.66 &3.596 &2.524 &1.666 &3.800 &0.0026 &0.0018 & -0.0062 & -0.0013 \\
 4 &64.37 &2.964 &3.950 &1.659 &4.372 & -0.0012 & -0.0022 &0.0008 &0.0016 \\
 5 &84.60 &2.576 &2.152 &3.009 &1.605 & -0.0068 & -0.0057 & -0.0081 & -0.0063 \\
\hline
\hline
\end{tabular*}
\label{tab:ndco_e_g}
\end{table*}

\begin{table*}
\caption{
Sm$^{3+}$ ion in SmCoO$_3$. Energy of three Kramers doublets originating from $^6H_{5/2}$ multiplet,
$\hat{g}$ and $\hat{\chi}^{vV}$ tensor components along the orthorhombic axes and $\omega$ direction.
Energy  $\varepsilon(0)$ is in meV, $\hat{\chi}^{vV}$ is units of $\mu_B$/T. }
\centering \setcellgapes{3.5pt} \makegapedcells
\begin{tabular*}{\textwidth}{@{\extracolsep{\fill}} cccccccccc}
\hline
\hline
doublet & $\varepsilon(0)$ &$g_{aa}$& $g_{bb}$& $g_{cc}$& $g_{\omega}$&$\chi^{vV}_{aa}$& $\chi^{vV}_{bb}$&$\chi^{vV}_{cc}$&$\chi^{vV}_{\omega}$ \\
\hline
 1 & 0.00 &0.671 &0.625 &0.322 &0.593 &0.0017 &0.0013 &0.0019 &0.0012 \\
 2 &29.63 &1.525 &0.963 &0.464 &0.398 & 0.0012 & 0.0017 &0.0012 &0.0011 \\
 3 &61.79 &1.141 &1.362 &0.067 &0.328 &0.0013 &0.0010 & 0.0008 & 0.0012 \\
\hline
\hline
\end{tabular*}
\label{tab:smco_e_g}
\end{table*}

\begin{table*}
\caption{
Dy$^{3+}$ ion in DyCoO$_3$. Energy of eight Kramers doublets originating from $^6H_{15/2}$ multiplet,
$\hat{g}$ and $\hat{\chi}^{vV}$ tensor components along the orthorhombic axes and $\omega$ direction.
Energy  $\varepsilon(0)$ is in meV, $\hat{\chi}^{vV}$ is units of $\mu_B$/T. }
\centering \setcellgapes{3.5pt} \makegapedcells
\begin{tabular*}{\textwidth}{@{\extracolsep{\fill}} cccccccccc}
\hline
\hline
doublet & $\varepsilon(0)$ &$g_{aa}$& $g_{bb}$& $g_{cc}$& $g_{\omega}$&$\chi^{vV}_{aa}$& $\chi^{vV}_{bb}$&$\chi^{vV}_{cc}$&$\chi^{vV}_{\omega}$ \\
\hline
 1 & 0.00 &8.52 &17.47 &0.03 &6.33 &0.0124 &0.0029 &0.0526 &0.0123 \\
 2 &29.75 &6.68 &14.90 &0.22 &5.82 & 0.0176 & 0.0049 &0.0210 &0.0174 \\
 3 &59.02 &7.42 &11.35 &0.38 &2.84 &0.0259 &0.0161 & 0.0490 & 0.0364 \\
 4 &80.47 &8.66 &6.92 &0.07 &1.34 & 0.0321 & 0.0434 &0.1143 &0.0553 \\
 5 &90.14 &18.65 &0.47 &0.02 &13.48 & 0.0298 & 0.0623 & -0.0087 & 0.1136 \\
 6 &96.11 &7.52 &4.40 &6.98 &7.28 &-0.0818 &-0.0677 &-0.0911 &-0.1653 \\
 7 &114.61 &9.89 &7.37 &0.30 &12.08 & -0.0122 & -0.0151 &0.2774 &-0.0164 \\
 8 &124.22 &10.58 &11.99 &2.62 &15.95 &-0.0227 &-0.0455 & -0.4132 & -0.0522 \\
\hline
\hline
\end{tabular*}
\label{tab:dyco_e_g}
\end{table*}

\begin{table*}
\caption{
Tb$^{3+}$ ion in TbCoO$_3$. Energy of five magnetic quasidoublets and three singlets
originating from $^7F_{6}$ multiplet,
$\hat{g}$ tensor components along the orthorhombic axes and $\omega$ direction,
in high field ($\sim 1$~T) limit. Energy  $\varepsilon(0)$ is in meV.}
\centering \setcellgapes{3.5pt} \makegapedcells
\begin{tabular*}{\tabws\textwidth}{@{\extracolsep{\fill}} cccccccccc}
\hline
\hline
singlet & $\varepsilon(0)$ &$g_{aa}$& $g_{bb}$& $g_{cc}$& $g_{\omega}$ \\
\hline
 1+2 & 0.000+0.002 &14.18 &10.71 &0 &2.40 \\
 3+4 &22.897+23.063 &1.59 &4.34 &0 &0.08 \\
 5 &33.264 &- &- &- &- \\
 6 &35.889 &- &- &- &- \\
 7+8 &41.325+41.334 &3.82 &0.39 &0 &1.76 \\
 9 &43.185 &- &- &- &- \\
 10+11 &43.59+44.491 &0 &$\sim0.04$ &0.31 &0 \\
 12+13 &68.797+68.824 &13.30 &5.97 &0 &14.66 \\
\hline
\hline
\end{tabular*}
\label{tab:tbco_e_g}
\end{table*}

\end{document}